# Statistical Mechanics in Collective Coordinates

S.F. Edwards

Polymer and Colloid Group

Cavendish Laboratory

Cambridge CB30HE-UK

Moshe Schwartz

Raymond and Beverly Sackler Faculty of Exact Sciences

School of Physics and Astronomy, Tel Aviv University,

Ramat Aviv, Tel Aviv 69978, Israel

## Abstract

We study the transformation of the statistical mechanics of N particles to the statistical mechanics of fields, that are the collective coordinates, describing the system. We give an explicit expression for the functional Fourier transform of the Jacobian of the transformation from particle to collective coordinate and derive the Fokker-Planck equation in terms of the collective coordinates. Simple approximations, leading to Debye-Huckel theory and to the hard sphere Percus-Yevick equation are discussed.

1 Introduction

Collective coordinates have provided a powerful insight into statistical problems notably in plasma physics [1], liquid helium [2], electrolytes [3] and the general theory of liquids [4, 5]. The use of collective coordinates raises several interesting problems, which we consider in this paper.

It is clear that when a classical system of identical particles is considered, all the physically relevant observables that depend only on the coordinates, are functionals of the density

$$\rho(\vec{r}) = \sum_\alpha \delta(\vec{r} - \vec{R}_\alpha), \qquad (1)$$

where $\vec{R}_\alpha$ are the coordinates of the particles (more general observables that depend also on momenta are functionals of the density and the current density). We consider a system of N particles enclosed in a cubic box of size L and periodic boundary conditions. The natural collective coordinates are the Fourier transforms, with $\vec{q} \neq 0$, of the density

$$\rho_q = \frac{1}{\sqrt{N}} \int \rho(\vec{r}) e^{i\vec{q}\cdot\vec{r}} d^3r, \quad \text{with} \quad q_i = \frac{2\pi n_i}{L}. \qquad (2)$$

An obvious problem in using the $\rho_q$'s, is that their number is infinite, while the number of degrees of freedom is 3N. This means that only 3N are independent and all the other collective coordinates can be written as functionals of the independent 3N. It is clear, however, that it is best to treat the collective coordinates in a symmetric way, keeping the infinite number of collective coordinates. We will show in the following, how this can be done by introducing the Jacobian of the transformation from particle to collective coordinates.

An additional difficulty comes with the introduction of hard cores. Experience has shown that approximation of a hard core by a soft core is inadequate for, as in all problems which are basically topological, the values of thermodynamic and transport coefficients is dominated by the softness and 'hardening' of a soft core always leads to false results. A major advance in the handling of the hard core was made by Percus and Yevick [4] and we discuss this in our context in section (3).

We will discuss first the Jacobian of the transformation from particle to collective coordinates and show that its functional Fourier transform has a rather simple form [6]. Next we show that the problem of an explicit form of the Jacobian of transformation can be avoided by deriving an exact Fokker-Planck equation in collective coordinates.

## 2 The Jacobian

An average of a quantity G that is a function of the density is given by

$$\langle G \rangle = \frac{\int \prod_\gamma dR_\gamma \, G\{\rho_\ell\} \exp\left[-\frac{1}{2kT} \sum_{\alpha,\beta} W(\vec{R}_\alpha - \vec{R}_\beta)\right]}{\int \prod_\gamma dR_\gamma \, \exp\left[-\frac{1}{2kT} \sum_{\alpha,\beta} W(\vec{R}_\alpha - \vec{R}_\beta)\right]}. \qquad (3)$$

Introducing $\delta$ functions, we can write the numerator $\mathbb{N}$ in (3) as well as the denominator in the form

$$\mathbb{N} = \int \mathcal{D}\rho(\vec{r}) \prod_\gamma dR_\gamma \prod_{\vec{q} \neq 0} \delta\left(\rho_q - \frac{1}{\sqrt{N}} \sum_\alpha e^{-i\vec{q}\cdot\vec{R}_\alpha}\right) G\{\rho_\ell\} \exp\left[-\frac{\bar{\rho}}{2kT} \sum W_\ell \rho_\ell \rho_{-\ell}\right], \qquad (4)$$

where $\mathcal{D}\rho$ denotes functional integration over general real functions $\rho(\vec{r})$ with fixed $\rho_0 = \sqrt{N}$, $W_\ell$ is the Fourier transform of the inter-particle potential, and $\bar{\rho}$ is the average density. (Note that $\bar{\rho}$ appears because of our definition of Fourier components of the density. Note also that since $\rho_0$ is fixed the inclusion of the self energy causes no problem.) Defining the Jacobian

$$J\{\rho\} = \int \prod_\gamma dR_\gamma \prod_{\vec{q} \neq 0} \delta\left(\rho_q - \frac{1}{\sqrt{N}} \sum_\alpha e^{-i\vec{q}\cdot\vec{R}_\alpha}\right), \qquad (5)$$

we see that $\mathbb{N}$ can be written as a functional integral over $\rho$

$$\mathbb{N} = \int \mathcal{D}\rho(\vec{r}) J\{\rho\} G\{\rho_\ell\} \exp\left[-\frac{\bar{\rho}}{2kT}\sum W_\ell \rho_\ell \rho_{-\ell}\right], \tag{6}$$

with a similar equation for the denominator.

Consider now the functional Fourier transform of the Jacobian

$$\mathcal{J}\{\varphi\} = \int \mathcal{D}\rho(\vec{r}) J\{\rho\} \exp\left[-i\sum_{\vec{q}\neq 0}\varphi_q \rho_{-q}\right], \tag{7}$$

using eq. (5), we find that

$$\mathcal{J}\{\varphi\} = \int \mathcal{D}\rho(\vec{r}) \prod_\gamma dR_\gamma \prod_{\vec{q}\neq 0}\delta\left(\rho_q - \frac{1}{\sqrt{N}}\sum_\alpha e^{-i\vec{q}\cdot\vec{R}_\alpha}\right)\exp\left[-i\sum_{\vec{q}\neq 0}\varphi_q\rho_{-q}\right] = \left[\Gamma\{\varphi\}\right]^N, \tag{8}$$

where

$$\Gamma\{\varphi\} = \int d\vec{R} \exp\left[-i\varphi(\vec{R})\right], \tag{9}$$

and

$$\varphi(\vec{r}) = -\frac{1}{\sqrt{N}}\sum_q \varphi_q e^{-i\vec{q}\cdot\vec{r}}. \tag{10}$$

Since N is extensively large, this can be written also in terms of a fugacity

$$\left[\Gamma\{\varphi\}\right]^N = \frac{N!}{2\pi i}\oint \frac{d\mu}{\mu^{N+1}}\exp\left[\mu\Gamma\{\varphi\}\right] \tag{11}$$

so we could use $\mathcal{J}\{\varphi,\mu\}$ instead of $\mathcal{J}\{\varphi,N\}$

$$\mathcal{J}\{\varphi,\mu\} = \exp\left[\mu\Gamma\{\varphi\}\right] \tag{12}$$

Thus

$$J\{\rho,\mu\} = \mathfrak{N}\int \mathcal{D}\varphi \exp\left[\mu\Gamma\{\varphi\}\right]\exp\left[i\sum\varphi_q\rho_{-q}\right], \tag{13}$$

where $\mathfrak{N}$ is the proper normalization and the sum now is over all q. Pertubation theory made by expanding (7) or (13) in $\varphi$ yields a simple approximation for $J\{\rho\}$. We will not dwell on this here because we will obtain those approximations from the dynamical equations to be considered in the following.

3. Dynamics

The dynamic problem is of obvious interest in its own right, but also offers a new way to approach the hard core. The original study of dynamics of Bohm & Pines produced plasma oscillation i.e. a property of $\ddot{\rho}$, but the problems we have been interested in are better expressed by noise methods of Langevin and Smoluchowski. Although current usage is to call these equations Fokker-Planck, they are not really F-P for this classic work on Brownian dynamic involves the velocities (the ancient paper of Chandrasekhar [7] is still a splendid introduction to this equation.) We want to use the $\rho$ and $\varphi$ directly and so base ourselves on the equation for $P(R\alpha...,t)$

$$\frac{\partial P}{\partial t} - \sum_\alpha D \frac{\partial}{\partial R_\alpha}\left(\frac{\partial}{\partial R_\alpha} + \frac{1}{kT}\frac{\partial W}{\partial R_\alpha}\right)P = 0 \qquad (14)$$

whose equilibrium solution, of course, is the Gibbs distribution $P \propto e^{-\frac{W}{kT}}$. Now introduce $P\{\rho\}$ via

$$P\{\rho,t\} = \int \prod_\alpha dR_\alpha \prod_{q\neq 0}\left[\rho_q - \frac{1}{\sqrt{N}}\sum_\beta e^{i\vec{q}\cdot\vec{R}_\beta}\right]P\{\vec{R}_\beta,t\} \qquad (15)$$

Therefore, multiply (14) by $\prod_{q\neq 0}\left[\rho_q - \frac{1}{\sqrt{N}}\sum_\beta e^{-i\vec{q}\cdot\vec{R}_\beta}\right]$ and integrate all over R to obtain

$$\frac{\partial P}{\partial t} - \frac{D}{\sqrt{N}}\sum_{k,j}(k\cdot j)\frac{\partial}{\partial \rho_k}\rho_{k+j}\frac{\partial}{\partial \rho_j}P + \sum_k Dk^2\frac{\partial}{\partial \rho_k}\rho_k P$$
$$-\sum \frac{\bar{\rho}D}{\sqrt{N}}(k\cdot j)\frac{\partial}{\partial \rho_k}\rho_{k-j}\rho_j\frac{W_j}{kT}P = 0 \qquad (16)$$

Similarly one can write this in Fourier variables $\varphi$

$$\frac{\partial P}{\partial t} + i \frac{D}{\sqrt{N}} \sum_{k,j} (k \cdot j) \varphi_{-k} \frac{\partial}{\partial \varphi_{-k-j}} \varphi_{-j} P + \sum_k Dk^2 \frac{\partial}{\partial \varphi_{-k}} \varphi_{-k} P \\ - i \sum \frac{\overline{\rho} D}{\sqrt{N}} (k \cdot j) \varphi_{-k} \frac{\partial}{\partial \varphi_{-k+j}} \frac{\partial}{\partial \varphi_{-j}} \frac{W_j}{kT} P = 0 \qquad (17)$$

and verify that the steady state solution is

$$P \propto \exp\left[ -\sum_k \varphi_k \left( \frac{kT}{4W} \right) \varphi_{-k} + N \log \int e^{-\varphi(R)} dR \right]. \qquad (18)$$

The main strength of the above equations is that it may be used to obtain directly hierarchies of equations for average quantities. Let $\mathbb{F}$ be a functional of the density that is not expected to diverge too strongly, like $\rho_q \rho_{-q}$, $\rho_{\ell_1} \rho_{\ell_2} \rho_{-\ell_1-\ell_2}$ etc. Suppose we are interested only in equilibrium properties. Then $\frac{\partial P}{\partial t} = 0$, we multiply eq. (16) by $\mathbb{F}$ and integrate by parts to obtain

$$\sum_k Dk^2 \left\langle \frac{\partial \mathbb{F}}{\partial \rho_k} \rho_k \right\rangle - \frac{D}{\sqrt{N}} \sum_{k,j} (k \cdot j) \left\langle \frac{\partial^2 \mathbb{F}}{\partial \rho_k \rho_j} \rho_{k+j} \right\rangle + \\ + \frac{\overline{\rho} D}{\sqrt{N}} \sum_{k,j} (k \cdot j) \frac{W_j}{kT} \left\langle \frac{\partial \mathbb{F}}{\partial \rho_k} \rho_{k-j} \rho_j \right\rangle = 0 \qquad (19)$$

It is clear that the $\rho_q$'s are dependent (being an infinite set of coordinates in contrast to the 3N components of the $R_\alpha$'s). Therefore, an infinite number of relations must hold among the $\rho_q$'s. This fact may lead to interesting sum rules.

A similar equation can be obtained for the dynamical problem in which a quantity $\mathbb{A}$ is measured at equilibrium, the system is allowed to evolve freely and then a quantity $\mathbb{B}$ is measured at a later time $t$. We stick at present to the equilibrium problem.

If a 'random phase' approximation is made by keeping the sum over k and j only those terms where one of the ρ's is $\rho_0 = \sqrt{N}$ (Note that derivatives are never with respect to $\rho_0$) the equation collapses to

$$\frac{\partial P}{\partial t} - \sum_k Dk^2 \frac{\partial}{\partial \rho_k}\left(\frac{\partial}{\partial \rho_{-k}} + \rho_k + \frac{\overline{\rho} W_k \rho_k}{kT}\right) P = 0 \qquad (20)$$

with the equilibrium solution

$$P_{eq} \propto \exp\left[-\frac{1}{2}\sum_k \left(1 + \frac{\overline{\rho} W_k}{kT}\right)\rho_k \rho_{-k}\right], \qquad (21)$$

that leads, for example, to the Debye-Huckel theory.

The simple RPA form given above can be used to derive the Percus-Yevick [4] equation for hard spheres. To do this we have to think of the hard sphere system as an ideal gas with a constraint that forces the pair distribution function to vanish inside the hard core radius. The pair distribution function is given by

$$g_2(\vec{r}_1 - \vec{r}_2) = \frac{1}{\overline{\rho}^2}\langle \rho(\vec{r}_1)\rho(\vec{r}_2) - \overline{\rho}\delta(\vec{r}_1 - \vec{r}_2)\rangle. \qquad (22)$$

Imposing the constraint is done by adding to the Hamiltonian a term $\frac{1}{2}\int d^3r d^3r' \lambda(\vec{r} - \vec{r}')\rho(\vec{r})\rho(\vec{r}')$ where the Lagrange multiplier makes $g_2$ vanish for $|\vec{r}_1 - \vec{r}_2| < R$, the hard core radius. The Lagrange multiplier must vanish outside the hard core radius because the constraint is just within that radius. Denoting by $\lambda_k$ the Fourier transform of $\lambda(\vec{r})$, we find that in eq. (20) $W_k$ is replaced by $\lambda_k$, so that

$$\langle \rho_k \rho_{-k}\rangle = \frac{1}{1 + \frac{\overline{\rho}\lambda_k}{kT}} \equiv \frac{1}{1 + \tilde{\lambda}_k}, \qquad (23)$$

where $\tilde{\lambda}_k$ is the Fourier transform of a function that vanishes outside the hard core radius. Our equations are thus $g_2(r) = 0$ for $r < R$ and $\tilde{\lambda}(r) = 0$ for $r \geq R$ [8], which are the Percus-Yevick equations for a system of hard spheres.

## 4 Acknowledgements